\documentclass[trackchanges]{aastex701}
\usepackage{amsmath}
\usepackage{booktabs}
\usepackage{CJKutf8}

\begin{document}
\begin{CJK*}{UTF8}{gbsn}

\title{Polarization of Fast Radio Bursts due to Propagation in Magnetized Plasma}

\author[orcid=0000-0001-9036-8543]{Wei-Yang Wang (王维扬)}
\affiliation{School of Astronomy and Space Science, University of Chinese Academy of Sciences, Beijing 100049, Peopleʼs Republic of China} 
\email[show]{wywang@ucas.ac.cn}  

\author[0000-0002-2552-7277]{Xiaohui Liu (刘小辉)}
\affiliation{State Key Laboratory of Radio Astronomy and Technology, National Astronomical Observatories, CAS, A20 Datun Road, Chaoyang District, Beijing, 100101, P. R. China}
\affiliation{School of Astronomy and Space Science, University of Chinese Academy of Sciences, Beijing 100049, People's Republic of China}
\email{liuxh@bao.ac.cn} 

\author[orcid=0009-0004-7249-8060]{Nan Xu(徐楠)}
\affiliation{The Hong Kong Institute for Astronomy and Astrophysics, The University of Hong Kong, Pokfulam Road, Hong Kong, China}
\affiliation{Department of Physics, The University of Hong Kong, Pokfulam Road, Hong Kong, China}
\email{nan.xu.astro@connect.hku.hk}

\author[orcid=0009-0000-6275-3452]{Songyu Shen(申嵩聿)}
\affiliation{School of Physics and Astronomy, Beijing Normal University, Beijing 100875, China}
\affiliation{Institute for Frontiers in Astronomy and Astrophysics, Beijing Normal University, Beijing 102206, China}
\email{syshen@mail.bnu.edu.cn}

%% Use the \collaboration command to identify collaborations. This command
%% takes an optional argument that is either a number or the word "all"
%% which tells the compiler how many of the authors above the command to
%% show. For example "\collaboration[all]{(DELVE Collaboration)}" wil include
%% all the authors above this command.
%%
%% Mark off the abstract in the ``abstract'' environment. 
\begin{abstract}

Polarization measurements of radio transients, particularly their frequency dependence, shed light on the magnetic environments of these sources.
We extend the analytical solution of radiative transfer of the polarization properties of such transients to both a cold magnetized plasmas beyond the high-frequency approximation and a relativistic plasma with synchrotron emission.
The Faraday mixing coefficients are enhanced makes Stokes parameters $Q$, $U$, and $V$ to exhibit a more rapid and complex frequency-dependent evolution.
A relativistic plasma medium can also generate both Faraday roataion and conversion with a smaller magnetic field.
The relativistic plasma is investigated as a persistent radio source (PRS) associated with FRB, and it is found that, although the observed PRS luminosity can be accounted for, an FRB can propagate out only under certain special parameter conditions.
We apply the model to fit the polarization profile of GPM J1839$-$10 and revisit some FRBs.
All sources with Faraday conversion require that the environment's magnetic field not exceed 100~G.
The magnetic field of FRB 20180301A and GPM J1839$-$10 are almost perpendicular to the light of sight for at least some times.

%This suggests even the PRS is associated with the FRB, the FRB does not traverse the PRS directly.

\end{abstract}

%% Keywords should appear after the \end{abstract} command. 
%% The AAS Journals now uses Unified Astronomy Thesaurus (UAT) concepts:
%% https://astrothesaurus.org
%% You will be asked to selected these concepts during the submission process
%% but this old "keyword" functionality is maintained in case authors want
%% to include these concepts in their preprints.
%%
%% You can use the \uat command to link your UAT concepts back its source.


%% From the front matter, we move on to the body of the paper.
%% Sections are demarcated by \section and \subsection, respectively.
%% Observe the use of the LaTeX \label
%% command after the \subsection to give a symbolic KEY to the
%% subsection for cross-referencing in a \ref command.
%% You can use LaTeX's \ref and \label commands to keep track of
%% cross-references to sections, equations, tables, and figures.
%% That way, if you change the order of any elements, LaTeX will
%% automatically renumber them.

\section{Introduction} 
\end{CJK*}
\linenumbers
Fast radio bursts (FRBs) manifest as millisecond-duration radio transients that can carry enormous energies.
Since the first discovery~\citep{2007Sci...318..777L}, FRB research has grown rapidly, with hundreds of sources now known~\citep{2021ApJS..257...59C,2026ApJS..283...34C}.
An FRB-like burst was detected from the Galactic magnetar SGR J1935+2154, providing the bridge between magnetars to at least some FRB progenitors~\citep{2020Natur.587...59B,2020Natur.587...54C}.
Nevertheless, the underlying radiation mechanisms and propagation effects that shape the observed bursts remain questions awaiting deeper investigation.

Polarization observations have emerged as a critical diagnostic tool, offering unique insights into both the radiation mechanisms, and crucially, the magneto-ionic environments.
While most FRBs exhibit high linear polarization, a subset displays significant circular polarization and complex polarization behaviors~\citep{2022Natur.611E..12X,2024arXiv240803313J,2024ApJ...972L..20N,2024ApJ...964..131S,2025ApJ...982..119B}.
Such observed polarization is invariably a hybrid product of the intrinsic radiation and extrinsic propagation effects.
Decoupling the intrinsic from the extrinsic is the prerequisite for transforming propagation effects from an obstacle to a diagnostic tool.

The most well-established propagation effect is Faraday rotation, quantified by the rotation measure (RM).
By combining with dispersion measure (DM), it provides a probe of the magnetic field strength and electron density in the source environment and intervening media. 
Some repeating FRBs exhibit stable RMs over long periods~\citep{2023ApJ...955..142Z}, while others display secular evolution or dramatic changes on timescales as short as days or even hours~\citep{2018Natur.553..182M,2020Natur.586..693L,2026MNRAS.546ag090O,2026ApJS..284...77W}, and even some sources have sign changes in RM~\citep{2023Sci...380..599A,2023MNRAS.526.3652K,2024ApJ...972L..20N}.

Beyond the standard Faraday rotation, more complex propagation effects, e.g., Faraday conversion, are invoked to account for more complex polarization observations.
Frequency-dependent linear and circular polarization degrees with intricate position angle (PA) variations, have been seen in sources like FRB 20180301A~\citep{2024MNRAS.534.2485U} and FRB 20201124A~\citep{2022Natur.611E..12X}, as well as two radio pulsars, the magnetar XTE J1810–197~\citep{2025Natur.637...43M} and the
long-period radio transient (LPT) GPM J1839–10~\citep{2025SciA...11P6351M}.
The ``generalized Faraday rotation'' (GFR) model offers a geometric interpretation of these phenomena by characterizing the trajectory of the polarization state on the Poincar\'e sphere~\citep{1997JPlPh..58..735M,1998PASA...15..211K}.
\cite{2025ApJ...988..164W} provides a physical description of propagation that is an analytical solution of radiative
transfer of the polarization properties.
However, in the cases of strongly magnetized cold plasmas and relativistic plasmas that may harbor non-thermal radiation, further interpretation is called for.

In this paper, we develop the radiative transfer model applicable to both a highly magnetized cold plasma and relativistic plasma with synchrotron emission, as two independent scenarios.
The paper is organized as
follows. 
The radiative transfer in the Stokes basis is discussed in Section~\ref{sec2}. We exhibit some
spectropolarimetric simulation results in Section~\ref{sec3}. 
Some
implications are discussed and consequences are summarized in Section~\ref{sec4}.

\section{Radiative Transfer with Emission in Magnetized Plasma}\label{sec2}

We take the transfer equation for FRB polarization as our point of departure.
This allows us to disentangle plasma and propagation effects.
The general solution of the transfer equation is universal across different plasma environments, with the plasma complexity entering solely through the corresponding terms in the equation.
Basically, we consider a homogeneous magnetized plasma. The transfer of the polarized emission in the Stokes basis can be described by \citep{1969SvA....13..396S}
\begin{equation}
\frac{d}{d s}\left(\begin{array}{l}
I \\
Q \\
U \\
V
\end{array}\right)=\left(\begin{array}{l}
\epsilon_I \\
\epsilon_Q \\
\epsilon_U \\
\epsilon_V
\end{array}\right)-\left(\begin{array}{cccc}
\eta_I & \eta_Q & \eta_U & \eta_V \\
\eta_Q & \eta_I & \rho_V & -\rho_U \\
\eta_U & -\rho_V & \eta_I & \rho_Q \\
\eta_V & \rho_U & -\rho_Q & \eta_I
\end{array}\right)\left(\begin{array}{l}
I \\
Q \\
U \\
V
\end{array}\right) ,
\label{eq:trans1}
\end{equation}
where $s$ is the path length, $\epsilon_S$, $\eta_S$ and $\rho_S$ are emissivity, absorption and Faraday mixing coefficients, respectively.

An intrinsic frame can be found in which all U-subscripted terms vanish. However, the observer's frame on Earth is not aligned with the intrinsic frame.
The observed Stokes vectors are referenced to the Earth's North, so that we introduce a rotation matrix:
\begin{equation}
\boldsymbol{R}(\chi_p)=\left(\begin{array}{cccc}
1 & 0 & 0 & 0 \\
0 & \cos 2 \chi_p & \sin 2 \chi_p & 0 \\
0 & -\sin 2 \chi_p & \cos 2 \chi_p & 0 \\
0 & 0 & 0 & 1
\end{array}\right),
\end{equation}
where $\chi_p$ is the angle between the intrinsic frame and the Earth's North.
The primed quantities refer to the observer's frame and symbols $\boldsymbol{\zeta}$, $\boldsymbol{\zeta_*}$, $\boldsymbol{q}$, $\boldsymbol{v}$, and $\boldsymbol{k}$ are defined as
\begin{equation}
\begin{aligned}
\boldsymbol{\zeta} &\equiv\left(\eta_Q' ,\eta_U',  \eta_{V}'\right), \\
\boldsymbol{\zeta_*} &\equiv\left(\rho_Q', \rho_U' , \rho_V'\right), \\
\boldsymbol{q} &\equiv\left(\eta_Q', \rho_Q'\right)\left[\kappa^2+\kappa_*^2\right]^{-1 / 2}, \\
\boldsymbol{u}& \equiv\left(\eta_U', \rho_U'\right)\left[\kappa^2+\kappa_*^2\right]^{-1 / 2}, \\
\boldsymbol{v}& \equiv\left(\eta_{V}', \rho_V'\right)\left[\kappa^2+\kappa_*^2\right]^{-1 / 2}, \\
\boldsymbol{k} &\equiv\left(\kappa, \kappa_*\right)\left[\kappa^2+\kappa_*^2\right]^{-1 / 2},
\end{aligned}
\end{equation}
and the two parameters $\kappa$ and $\kappa_*$ denote 
\begin{equation}
\begin{aligned}
&\kappa =\frac{1}{\sqrt2}\left\{\left[\left(\zeta^2-\zeta_*^2\right)^2+4\left(\boldsymbol{\zeta}\cdot \boldsymbol{\zeta_*}\right)^2\right]^{1 / 2}+\zeta^2-\zeta_*^2\right\} ^{1 / 2}, \\
&\kappa_* =\frac{1}{\sqrt2}\left\{\left[\left(\zeta_*^2-\zeta^2\right)^2+4\left(\boldsymbol{\zeta}\cdot \boldsymbol{\zeta_*}\right)^2\right]^{1 / 2}+\zeta_*^2-\zeta^2\right\}^{1 / 2}.
\end{aligned}
\end{equation}

The case of no absorption and pure Faraday effects needs to be treated with the general solution or the continuous limit.
In our prior work, we ignore the emissivities because the brightness temperature of FRB is much higher than the temperature of the plasma medium~\citep{2025ApJ...988..164W}.
The solution thus obtained works for cold and uniformly magnetized plasma, in which $\bar\gamma =kT/(m_e c^2)\ll1$.
In this work, we extend the solution of the transfer equation by invoking emissivity.
With the emissivities, the Stokes parameters now converge to finite value at infinity, $\vec{X}_p=(I_\infty,\,Q_\infty,\,U_\infty,\,V_\infty)^{\intercal}$, instead of naively being set to zero~\citep{1975ApJ...196..125P,1977ApJ...214..522J}.
Therefore, the extended solution of the transfer equation is\footnote{$(I_0,\,Q_0,\,U_0,\,V_0)$ is defined in the observer's frame, not the intrinsic plasma frame adopted in previous work.}
\begin{equation}
\begin{aligned}
\vec{S'}&=\vec{X}_p+\exp(-\tau)\\
&\times\left[\left(\begin{array}{cccc}
\frac{1}{2}\left(1+q^2+u^2+v^2\right) & -\boldsymbol{u} \times \boldsymbol{v} &-\boldsymbol{v} \times \boldsymbol{q} & -\boldsymbol{q} \times \boldsymbol{u} \\
\boldsymbol{u} \times \boldsymbol{v} & \frac{1}{2}\left(1+q^2-u^2-v^2\right) & \boldsymbol{q} \cdot \boldsymbol{u} &\boldsymbol{q} \cdot \boldsymbol{v} \\
\boldsymbol{v} \times \boldsymbol{q} & \boldsymbol{q} \cdot \boldsymbol{u}  & \frac{1}{2}\left(1-q^2+u^2-v^2\right) & \boldsymbol{u} \cdot \boldsymbol{v}  \\
\boldsymbol{q} \times \boldsymbol{u} &\boldsymbol{q} \cdot \boldsymbol{v} & \boldsymbol{u} \cdot \boldsymbol{v}  & \frac{1}{2}\left(1-q^2-u^2+v^2\right)
\end{array} \right)\, \cosh (\kappa s)\right. \\
& -\left(\begin{array}{cccc}
0 &\boldsymbol{q} \cdot \boldsymbol{k} & \boldsymbol{u} \cdot \boldsymbol{k}  & \boldsymbol{v} \cdot \boldsymbol{k} \\
\boldsymbol{q} \cdot \boldsymbol{k} & 0 & -\boldsymbol{v} \times \boldsymbol{k} & \boldsymbol{u} \times \boldsymbol{k}  \\
\boldsymbol{u} \cdot \boldsymbol{k}  & \boldsymbol{v} \times \boldsymbol{k} & 0 & -\boldsymbol{q} \times \boldsymbol{k} \\
\boldsymbol{v} \cdot \boldsymbol{k} & -\boldsymbol{u} \times \boldsymbol{k}  &\boldsymbol{q} \times \boldsymbol{k} & 0
\end{array}\right) \sinh (\kappa s) \\
& +\left(\begin{array}{cccc}
\frac{1}{2}\left(1-q^2-u^2-v^2\right) & -\boldsymbol{v} \times \boldsymbol{u} & -\boldsymbol{q} \times \boldsymbol{v} & -\boldsymbol{u} \times \boldsymbol{q} \\
\boldsymbol{v} \times \boldsymbol{u} & \frac{1}{2}\left(1-q^2+u^2+v^2\right) & -\boldsymbol{q} \cdot \boldsymbol{u} & -\boldsymbol{q} \cdot \boldsymbol{v} \\
\boldsymbol{q} \times \boldsymbol{v} & -\boldsymbol{q} \cdot \boldsymbol{u} & \frac{1}{2}\left(1+q^2-u^2+v^2\right) & -\boldsymbol{u} \cdot \boldsymbol{v} \\
\boldsymbol{u} \times \boldsymbol{q} & -\boldsymbol{q} \cdot \boldsymbol{v} & -\boldsymbol{u} \cdot \boldsymbol{v} & \frac{1}{2}\left(1+q^2+u^2-v^2\right)
\end{array}\right) \cos \left(\kappa_* s\right) \\
& \left.-\left(
\begin{array}{cccc}
0 &\boldsymbol{q} \times \boldsymbol{k} & \boldsymbol{u} \times \boldsymbol{k} & \boldsymbol{v} \times \boldsymbol{k} \\
\boldsymbol{q} \times \boldsymbol{k} & 0 & \boldsymbol{v} \cdot \boldsymbol{k} & -\boldsymbol{u} \cdot \boldsymbol{k} \\
\boldsymbol{u} \times \boldsymbol{k} & -\boldsymbol{v} \cdot \boldsymbol{k} & 0 &\boldsymbol{q} \cdot \boldsymbol{k} \\
\boldsymbol{v} \times \boldsymbol{k} & \boldsymbol{u} \cdot \boldsymbol{k} & -\boldsymbol{q} \cdot \boldsymbol{k} & 0
\end{array}\right) \sin \left(\kappa_* s\right)\right]\left(\begin{array}{l}
I_0-I_\infty \\
Q_0-Q_\infty \\
U_0-U_\infty \\
V_0-V_\infty
\end{array}\right),\\
&
\end{aligned}
\label{eq:solution}
\end{equation}
where $\tau=\eta_Is$ is the optical depth, the superscript ``0'' for $s = 0$, and 
%\begin{strip}
\begin{equation}
\begin{aligned}
I_\infty&=\frac{\eta'_I}{(\eta_I'^2-\kappa^2)(\eta_I'^2+\kappa_*^2)}\{\left(\eta_I'^2+\zeta_*^2\right)\epsilon'_I
-\left[\eta_Q'\eta_I'+\left(\vec{\zeta}\cdot\vec{\zeta_*}\right) \frac{\rho'_Q}{\eta_I'}+\rho_V'\eta_U'-\eta_{\nu}'\rho_U'\right] \epsilon_Q'\\
&-\left[\eta_U'\eta_I'+\left(\vec{\zeta}\cdot\vec{\zeta_*}\right) \frac{\rho'_U}{\eta_I'}+\rho_Q'\eta_{\nu}'-\eta_Q'\rho_V'\right] \epsilon_U'
-\left[\eta_{\nu}'\eta_I'+\left(\vec{\zeta}\cdot\vec{\zeta_*}\right) \frac{\rho'_V}{\eta_I'}+\rho_U'\eta_Q'-\eta_U'\rho_Q'\right] \epsilon_V'
\},\\
Q_\infty&=\frac{\eta'_I}{(\eta_I'^2-\kappa^2)(\eta_I'^2+\kappa_*^2)}
\{\left[-\eta_Q'\eta_I'-\left(\vec{\zeta}\cdot\vec{\zeta_*}\right) \frac{\rho'_Q}{\eta_I'}+\rho_V'\eta_U'-\eta_{\nu}'\rho_U'\right] \epsilon_I'
+(\eta_I'^2+\rho_Q'^2-\eta_U'^2-\eta_{\nu}'^2)\epsilon_Q'\\
&+\left[-\rho_V'\eta_I'+\left(\vec{\zeta}\cdot\vec{\zeta_*}\right) \frac{\eta'_V}{\eta_I'}+\rho_Q'\rho_U'+\eta_Q'\eta_U'\right] \epsilon_U'
+\left[\rho_U'\eta_I'-\left(\vec{\zeta}\cdot\vec{\zeta_*}\right) \frac{\eta'_U}{\eta_I'}+\eta_{\nu}'\eta_Q'+\rho_V'\rho_Q'\right] \epsilon_V'
\},\\
U_\infty&=\frac{\eta'_I}{(\eta_I'^2-\kappa^2)(\eta_I'^2+\kappa_*^2)}
\{\left[-\eta_U'\eta_I'-\left(\vec{\zeta}\cdot\vec{\zeta_*}\right) \frac{\rho'_U}{\eta_I'}+\rho_Q'\eta_{\nu}'-\eta_Q'\rho_V'\right] \epsilon_I'
+\left[\rho_V'\eta_I'-\left(\vec{\zeta}\cdot\vec{\zeta_*}\right) \frac{\eta'_V}{\eta_I'}+\rho_Q'\rho_U'+\eta_Q'\eta_U'\right] \epsilon_Q'\\
&+(\eta_I'^2-\eta_Q'^2+\rho_U'^2-\eta_{\nu}'^2)\epsilon_U'
+\left[-\rho_Q'\eta_I'+\left(\vec{\zeta}\cdot\vec{\zeta_*}\right) \frac{\eta'_Q}{\eta_I'}+\eta_{\nu}'\eta_U'+\rho_V'\rho_U'\right] \epsilon_V'
\},\\
V_\infty&=\frac{\eta'_I}{(\eta_I'^2-\kappa^2)(\eta_I'^2+\kappa_*^2)}
\{\left[-\eta_{\nu}'\eta_I'-\left(\vec{\zeta}\cdot\vec{\zeta_*}\right) \frac{\rho'_V}{\eta_I'}+\rho_U'\eta_Q'-\eta_U'\rho_Q'\right] \epsilon_I'
+\left[-\rho_U'\eta_I'+\left(\vec{\zeta}\cdot\vec{\zeta_*}\right) \frac{\eta'_U}{\eta_I'}+\rho_Q'\rho_V'+\eta_Q'\eta_{\nu}'\right] \epsilon_Q'\\
&+\left[\rho_Q'\eta_I'-\left(\vec{\zeta}\cdot\vec{\zeta_*}\right) \frac{\eta'_Q}{\eta_I'}+\eta_{\nu}'\eta_U'+\rho_V'\rho_U'\right]\epsilon_U'
+(\eta_I'^2-\eta_Q'^2-\eta_U'^2+\rho_V'^2)\epsilon_V'
\}.\\
\end{aligned}
\end{equation}
%\end{strip}

\section{Spectro-Polarimetric Properties}\label{sec3}

\subsection{Cold and Highly Magnetized Plasma}\label{sec3.1}

\begin{figure*}[ht!]
\plotone{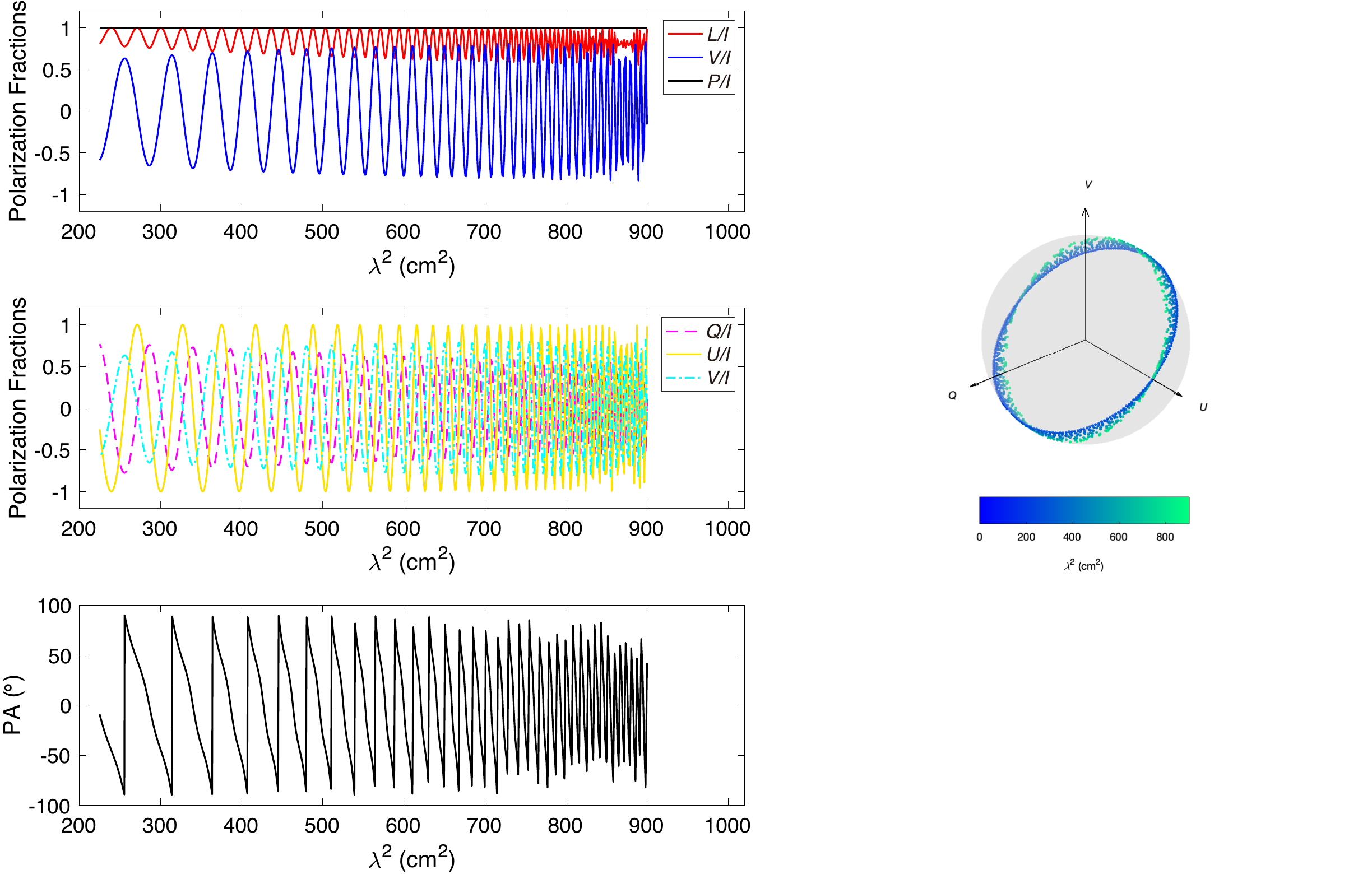}
\caption{Simulated spectropolarization profiles. (a) Top panel: linear, circular, and total polarization fractions; Middle panel: $Q/I$, $U/I$, and $V/I$. Bottom panel:
polarization angle. (b) Poincar\'e sphere representation of the spectropolarimetric properties. The parameters are adopted as $\theta_B= 105^\circ$, $n_0L =10^{13}\,cm^{-2}$, $B = 3\times 10^2$\,G, $Q_0/I_0 = 0$, $U_0/I_0 = 1$, and $V_0/I_0 = 0$.
}
\label{fig:prop1}
\end{figure*}

Consider the FRB waves traveling through a cold magnetized plasmas beyond the high-frequency approximation.
The highly magnetized plasma medium here implies that the cyclotron frequency is not consistent with $\omega_B\ll\omega$.
Our concern is the plasma medium, as opposed to a dense one.
The dominant absorption is free-free absorption, with its absorption coefficient being smaller than the mixing Faraday coefficients.

The mixing Faraday coefficients are determined by the differential phase delay between modes with different dispersions and polarizations.
The plasma is thought to be thermally distributed where $\bar\gamma=kT/(mc^2)\ll1$, where $m$ is the mass of electron.
We derive the dielectric tensor for our specified plasma parameters:
\begin{equation}
\epsilon^\mu_\nu=\delta^\mu_\nu+\frac{4\pi c}{\omega^2}\alpha^\mu_\nu,
\end{equation}
where the response tensor is given by \citep{1958PhDT........18T}
\begin{equation}
\begin{gathered}
\alpha_\nu^\mu(k)=\frac{i\omega_p^2 \omega}{4\pi c \bar\gamma^{2}K_2(\bar\gamma^{-1})} \int_0^{\infty} d \xi\left[t_\nu^\mu \frac{K_2(r)}{r^2}-R^\mu \bar{R}_\nu \frac{K_3(r)}{r^3}\right], \\
t_\nu^\mu=\left(\begin{array}{cc}
\cos ^2 \theta_B \cos \omega_B \xi+\sin ^2 \theta_B & \eta \cos \theta_B \sin \omega_B \xi \\
-\eta \cos \theta_B \sin \omega_B \xi & \cos \omega_B \xi
\end{array}\right), \\
R^\mu=\frac{\omega \sin \theta_B}{\omega_B}\left[\cos \theta_B\left(\sin \omega_B \xi-\omega_B \xi\right),-\eta\left(1-\cos \omega_B \xi\right)\right], \\
\bar{R}_\nu=\frac{\omega \sin \theta_B}{\omega_B}\left[\cos \theta_B\left(\sin \omega_B \xi-\omega_B \xi\right), \eta\left(1-\cos \omega_B \xi\right)\right], \\
r=\left[\frac{1}{\bar\gamma^2}-2 i \frac{\omega \xi}{\bar\gamma} +\frac{\omega^2 \sin ^2 \theta_B}{\omega_B^2}\left(2-\omega_B^2 \xi^2-2 \cos \omega_B \xi\right)\right]^{1 / 2},
\end{gathered}
\label{eq:response}
\end{equation}
where $\eta$ is the sign of the charge, $\omega_B$ is the cyclotron frequency and $\xi$ is the time.
%Note that the only significant contribution to the integral over proper time is from the range $\omega_B\xi\ll1$.
The first terms of Equation (\ref{eq:response}) can be written as \citep{2008ApJ...688..695S}
\begin{equation}
\begin{gathered}
r^2=r_0^2+\delta r^2, \quad r_0^2=\frac{1}{\bar\gamma^2}-2 i \frac{\omega \xi}{\bar\gamma}, \quad \delta r^2=-\frac{\sin ^2 \theta_B}{12} \omega^2 \omega_B^2\xi^4.
\end{gathered}
\label{eq:r0}
\end{equation}
We introduce the following integral forms of the modified Bessel function of the second kind:
\begin{equation}
\begin{aligned}
& \int_0^{\infty} d x\left[x^n \frac{K_2\left(\sqrt{\bar\gamma^{-2}-2 i \bar\gamma^{-1} x}\right)}{\bar\gamma^{-2}-2 i \bar\gamma ^{-1}x}\right]=n!i^{n+1} \frac{K_{n-1}(\bar\gamma^{-1})}{\bar\gamma^{-2}}, \\
& \int_0^{\infty} d x\left[x^n \frac{K_3\left(\sqrt{\bar\gamma^{-2}-2 i \bar\gamma^{-1} x}\right)}{\left(\bar\gamma^{-2}-2 i \bar\gamma^{-1} x\right)^{3 / 2}}\right]=n!i^{n+1} \frac{K_{n-2}(\bar\gamma^{-1})}{\bar\gamma^{-3}}.
\end{aligned}
\label{eq:integral}
\end{equation}
The terms related to Faraday rotation and Faraday conversion are given by
\begin{equation}
\begin{gathered}
\rho_V=-i\frac{4\pi}{\omega}\alpha_2^1, \quad \rho_Q=-\frac{2\pi}{\omega}(\alpha_1^1-\alpha_2^2).
\end{gathered}
\end{equation}
Combining Eqs. (\ref{eq:r0}) and (\ref{eq:integral}), the terms in which $K_3(x)$ appears as the integrand can be neglected.
The the terms related to Faraday rotation and Faraday conversion are then given by
\begin{equation}
\begin{aligned}
\rho_V&=-\frac{\omega_p^2\eta\cos\theta_B}{c\omega K_2(\bar\gamma^{-1})}\sum_{n=0}^{\infty}K_{2n}(\bar\gamma^{-1})\left(\frac{\omega_B}{\omega}\right)^{2n+1},\\
\rho_Q&=-\frac{\omega_p^2\sin^2\theta_B}{2c\omega K_2(\bar\gamma^{-1})}\sum_{n=1}^{\infty}K_{2n-1}(\bar\gamma^{-1})\left(\frac{\omega_B}{\omega}\right)^{2n}.
\end{aligned}
\label{eq:rho}
\end{equation}
The asymptotic expansion of the modified Bessel function is given by
\begin{equation}
K_n(x) \approx\sqrt{\frac{\pi}{2 n}}\left(\frac{e x}{2 n}\right)^{-n}\left[1+\frac{4 n^2-1}{8 nx}+\frac{\left(4 n^2-1\right)\left(4 n^2-9\right)}{2!(8 nx)^2}+\cdots\right],
\end{equation}
whose leading-order term will be used in the discussion of the convergence of the summation.

To examine the convergence, we consider the ratio of successive terms:
\begin{equation}
%\begin{aligned}
%&\begin{aligned}
\left|\frac{a_{n+1}}{a_n}\right|\sim \left(\frac{4 \bar{\gamma}\omega_B}{e\omega}\right)^2 \left(\frac{n+1}{n}\right)^{2 n+2}
%\end{aligned}\\
\sim\left(4 \bar{\gamma}\frac{\omega_B}{\omega}\right)^2.
%\end{aligned}
\end{equation}
We therefore obtain the convergence condition is $\omega_B/\omega<(4\bar\gamma)^{-1}$.
The Faraday rotation and Faraday conversion coefficients can be then approximated as
\begin{equation}
\begin{aligned}
\rho_V&=-\frac{\omega_p^2\omega_B\eta\cos\theta_B}{c(\omega^2-\omega_B^2)},\\
\rho_Q&=-\frac{\omega_p^2\omega_B^2\sin^2\theta_B}{2c\omega(\omega^2-\omega_B^2)}.
\end{aligned}
\label{eq:terms}
\end{equation}

It is assumed that the intrinsic FRB emissions are highly polarized, and their polarizations are independent with frequency.
When the magnetic field is aligned with the line of sight, the two eigenmodes experience different refractive indices, leading to a progressive rotation of the linear polarization plane, this is, Faraday rotation.
The PA change due to the Faraday rotation is $\Delta \mathrm{PA}=\mathrm{RM}\lambda^2$ when $\omega_B\ll\omega$.
In the strong-field regime, however, the simple linear scaling between PA and $\mathrm{RM}\lambda^2$ breaks down.
If we retain the conventional definition of RM, the PA change due to the Faraday rotation is then governed by
\begin{equation}
\Delta\mathrm{PA}=\mathrm{RM}\lambda^2\lambda_B^2/(\lambda_B^2-\lambda^2),  
\end{equation}
where $\lambda_B=2\pi c/\omega_B$.

Consider a cold plasma ($\eta=-1$) where the magnetic field is oriented obliquely with respect to the line of sight (LOS).
In such a strongly magnetized plasma, the polarization evolution of propagating radiation is governed not only by the Faraday rotation but also by Faraday conversion.
For simplicity, we take $\chi_p=0$ in the following simulations.
For our present interest, we consider the case that Faraday rotation and Faraday conversion are comparable.
The Faraday mixing coefficients are independent of temperature if $\bar\gamma\ll1$, so we take $T=1\,\rm K$.
The plasma medium will not give rise to a huge DM; at least it should be smaller than the DM observed in most FRBs, and its corresponding $n_0L<10^{21}\,\rm{cm}^{-2}$.
The plasma medium may correspond to an observable RM, so we set $n_0L=10^{13}\,\rm{cm}^{-2}$.
We take $B = 3 \times 10^2\,\rm{G}$ and $\theta_B = 105^\circ$ to compare with the results of previous work~\citep{2025ApJ...988..164W}, which only considered the first-order terms.
The simulated spectrum is shown in Figure \ref{fig:prop1}.
By invoking the higher order terms, the mixing Faraday coefficients are enhanced, and the polarization spectrum aries more rapidly with observing frequency as well.

\begin{figure*}[ht!]
\plotone{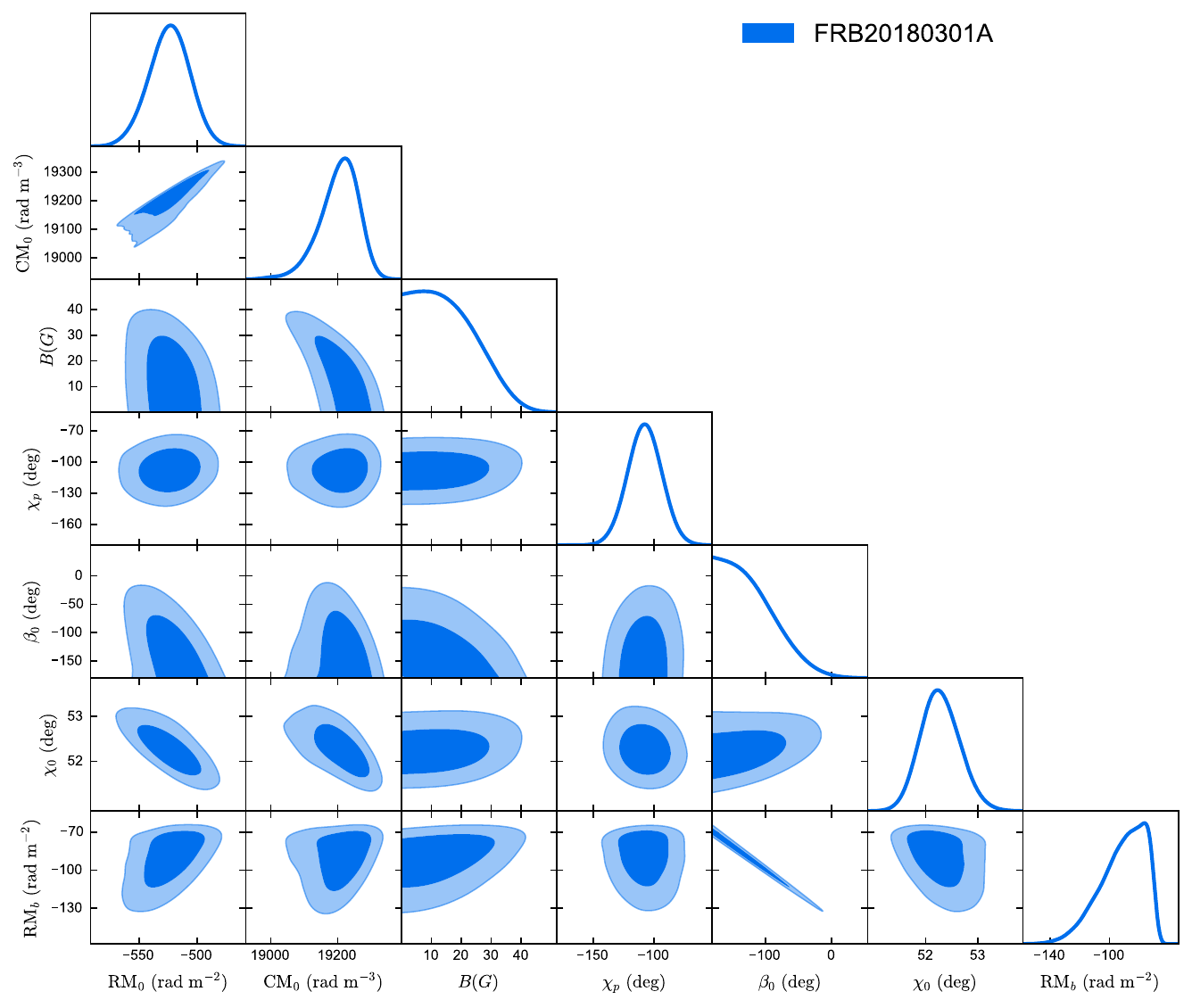}
\caption{The 1D and 2D marginalized probability distributions at 68.3\% and 95.4\% confidence levels for the cold plasma scenario for FRB 20180301A.
\label{fig:180301}}
\end{figure*}

\begin{figure*}[ht!]
\plotone{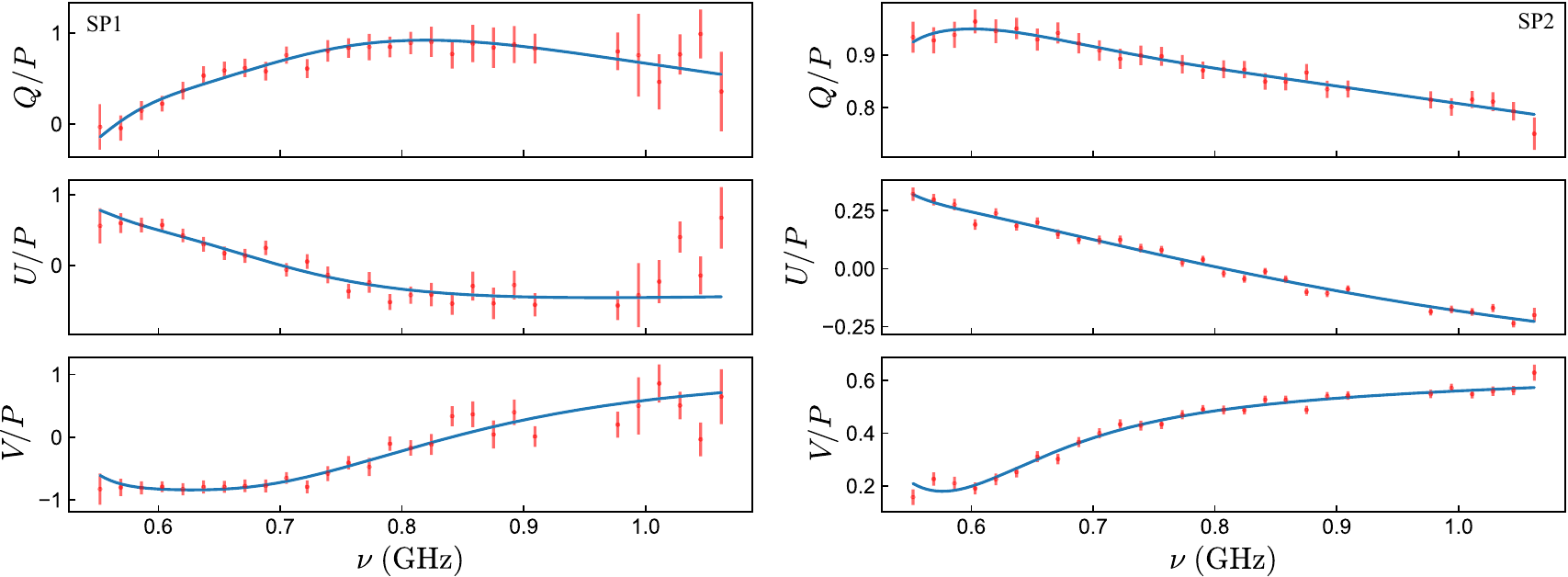}
\caption{The best fit of the model for GPM J1839$-$10.
\label{fig:sp1sp2}}
\end{figure*}

\begin{figure*}[ht!]
\plotone{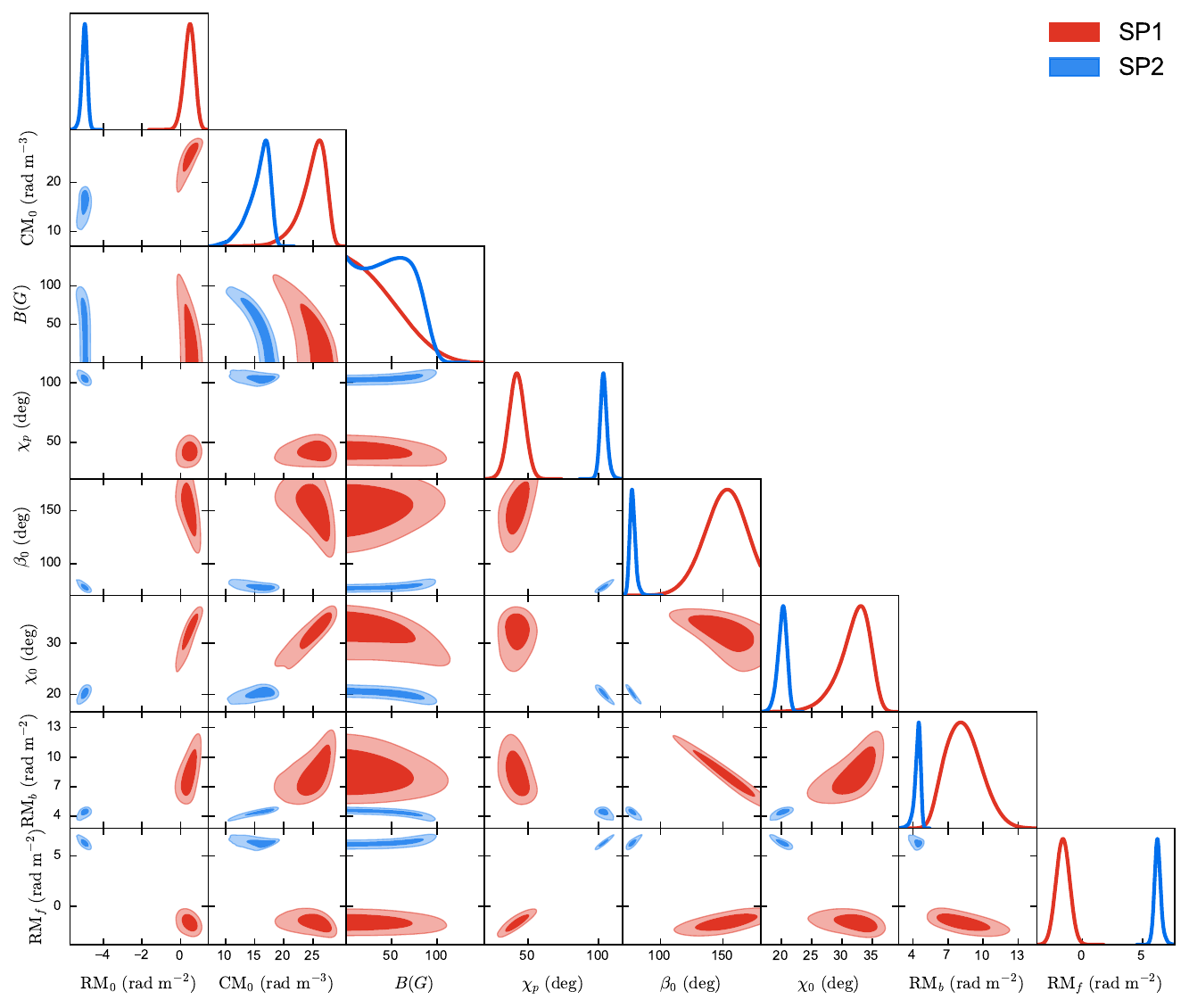}
\caption{Same as Figure~\ref{fig:180301} but for SP1 and SP2 of GPM J1839$-$10.
\label{fig:GPM}}
\end{figure*}

Bayesian inference is performed using \texttt{emcee} with uniform priors to sample from the posterior distributions~\citep{2013ascl.soft03002F}.
This method was adopted in the previous work, where we developed the \texttt{mixF} code and used it to fit the polarization properties of some FRBs~\citep{2025ApJ...988..164W}.
The initial Stokes parameters are converted as
\begin{equation}
Q_0=I_0\cos2\beta_0\cos2\chi_0,\, U_0=I_0\sin2\beta_0\cos2\chi_0,\,
V_0=I_0\sin2\chi_0.
\end{equation}
Since the spatial origin of the Faraday mixing plasma medium remains unknown, both a foreground and a background RM layer are introduced.
While there are some degeneracy between $B$, $n_0L$, and $\theta_B$.
Here we introduce $\mathrm{RM}_0$ and $\mathrm{CM}_0$ which can reduce the number of parameters.
We can recombine the mixing Faraday coefficients to obtain these two parameters:
\begin{equation} \label{eq:Bn0L}
\begin{aligned}
&B \frac{\sin^2 \theta_B}{\cos \theta_B} = \frac{\mathrm{CM}_0}{\mathrm{RM}_0} \times \frac{2 \pi m_e c^2}{e},\\
&n_0 L \tan^{-2} \theta_B = \frac{\mathrm{RM}_0^2}{\mathrm{CM}_0} \times \frac{m_e c^2}{e^2}.
\end{aligned}
\end{equation}
The introduction of $\rm{RM}_0$ and $\rm{CM_0}$ improves the convergence of the fit, thereby avoiding the poor convergence caused by the large disparity in magnitude between $n_0L$ and $B$. 

We adopt wide-uniform priors on all model parameters. $\mathrm{CM}_0$ should be a positive value, so it is restricted to  $[ 0,\,10^{5} ]\,\mathrm{rad~ m^{-3}}$.
Since the requirement that the observing frequency should lie above the cut-off frequency of both right- and left-hand modes, a uniform prior of $\left[ 0,190 \right]$ G\footnote{The magnetic field of 190 G corresponds to a cyclotron frequency of 500 MHz.} is adopted. Periodic angular parameters are restricted to a single fundamental period to avoid multiple representations of physically equivalent configurations.
The marginalized posterior densities are plotted using the GetDist Python package~\citep{Lewis:2019xzd}.

The modeled parameters of FRB 20180301A are listed in Table~\ref{tab:180301}.
The foreground RM of is consistent with the value fitted by the GFR model~\citep{2024MNRAS.534.2485U}.
The magnetic field lies almost orthogonal to the LOS, since the amplitude of the circular polarization variations in FRB 20180301A is comparable to the total intensity.
Although the plasma medium provides only a small RM of $-1.7~\text{rad~m}^{-2}$, this tiny amount of Faraday rotation, in conjunction with Faraday conversion, is still sufficient to generate oscillatory patterns in the polarization spectra across wavelength.
Notably, the source exhibits dramatic RM variations over timescales of several years~\citep{2020Natur.586..693L,2023MNRAS.526.3652K}, suggesting a changing or structured magnetized environment.
This turbulent environment also raises the possibility that, at certain epochs, the magnetic field within the circumsource medium may become oriented perpendicular to the LOS.

We also fit to the Burst 926 reported by~\cite{2022Natur.611E..12X}.
The results are indistinguishable from those of our previous work, which can be attributed to the magnetic field strength of $<100$~G, a regime in which the strong-field correction to the Faraday effect is negligible.

The LPT GPM J1839$-$10 has a high linear polarization and exhibits a frequency-dependent evolution of its Stokes parameters~\citep{2025SciA...11P6351M}.
We fit the observed Stokes parameters using the model.
While frequency-dependent circular polarization is seen in all three single pulses, the oscillation is less evident in SP3, and we therefore focus on SP1 and SP2 in what follows.
The best fitting polarization profile are shown in Figure~\ref{fig:sp1sp2}.
The marginalized posterior densities for the model are plotted in Figure~\ref{fig:GPM} and fitted parameters are listed in Table~\ref{tab:180301}.
Similar with FRB 20180301A, the magnetic field of the GPM source is oriented roughly perpendicular to the LOS, a geometry required for producing significant Faraday conversion.
Note that a known RM of $531.63~\mathrm{rad~m^{-2}}$ was adopted to correct the the pulses for Faraday rotation.
The true foreground RM of a pulse is the sum of the fixed RM and the $\mathrm{RM}_f$ derived from the model fit.
Furthermore, the two GPM bursts exhibit intrinsic circular polarization fractions of 90\% and 65\%, respectively.
Such high degrees of circular polarization have been widely seen in LPTs~\citep{2025A&A...699A.341B,2025ApJ...988L..29D,2025SciA...11P6351M}.

\begin{table}
\centering
\caption{68.3\% credible intervals for the parameters of the cold plasma model.}
\begin{tabular}{ccccccccc}
\toprule
Source & $\log_{10}(B/\rm{G})$ & $\theta_B$ (deg) & $\log_{10}(n_0L/\rm{cm^{-2}})$ & $\beta_0$ (deg) & $\chi_0$ (deg) & $\chi_p$ (deg) & $\rm RM_{\it f}\,(rad\,m^{-2})$ & $\rm RM_{\it b}\,(rad\,m^{-2})$ 
\\
\midrule
FRB 20180301A & $1.16^{+0.26}_{-0.49}$ & $90.23^{+0.09}_{-0.20}$ & $16.81^{+0.29}_{-0.97}$ & $-129.05^{+48.31}_{-35.18}$ & $52.26^{+0.39}_{-0.36}$ & $-107.70^{+14.00}_{-13.88}$ & $27.52\pm5.30$ & $-88.63^{+13.30}_{-18.31}$\\
GPM SP1 & $1.52_{-0.52}^{+0.31}$ & $89.73^{+0.27}_{-0.19}$ & $13.2^{+0.4}_{-1.1}$ & $151.60^{+15.05}_{-16.86}$ & $32.36^{+2.09}_{-2.97}$ & $41.95\pm0.55$ & $-1.62\pm0.56$ & $8.27^{+1.62}_{-1.44}$\\
GPM SP2 & $1.66^{+0.22}_{-0.50}$ & $98.2^{+2.8}_{-7.8}$ & $12.28^{+0.03}_{-0.42}$ & $77.58^{+2.39}_{-1.98}$ & $20.19^{+0.68}_{-0.80}$ & $103.46^{+2.46}_{-2.25}$ & $6.27^{+0.27}_{-0.25}$ & $4.44^{+0.20}_{-0.29}$\\
\bottomrule
\end{tabular}
\label{tab:180301}
\end{table}

Although Table 1 shows that the magnetic field is nearly perpendicular to the LOS, the plasma medium itself can still contribute to a RM, which means that Faraday rotation remains non-negligible compared to conversion.
As can be seen from Eq. (\ref{eq:rho}), when $\omega_B$ approaches $\omega$, the effect induced by the strong magnetic fields synchronously for both the rotation and conversion coefficients, and does not alter their ratio.
When the magnetic field strengthens, an increase in Faraday conversion is inevitably accompanied by a corresponding enhancement in Faraday rotation.

\subsection{Nonthermal Relativistic Plasma}\label{sec3.2}

Some FRBs are associated with persistent radio sources (PRSs)~\citep{2017Natur.541...58C,2022Natur.606..873N,2024Natur.632.1014B}.
These are widely interpreted as synchrotron nebulae powered by young magnetars or pulsar wind nebulae, and thus represent natural candidates for magnetized relativistic plasmas.
FRBs are most likely propagate through the plasmas, and their polarization would be reshaped.
We investigate whether PRSs, as magnetized relativistic plasmas, can simultaneously provide Faraday rotation and conversion, and to assess their role in shaping the observed polarization of FRBs.

\begin{figure*}[ht!]
\plotone{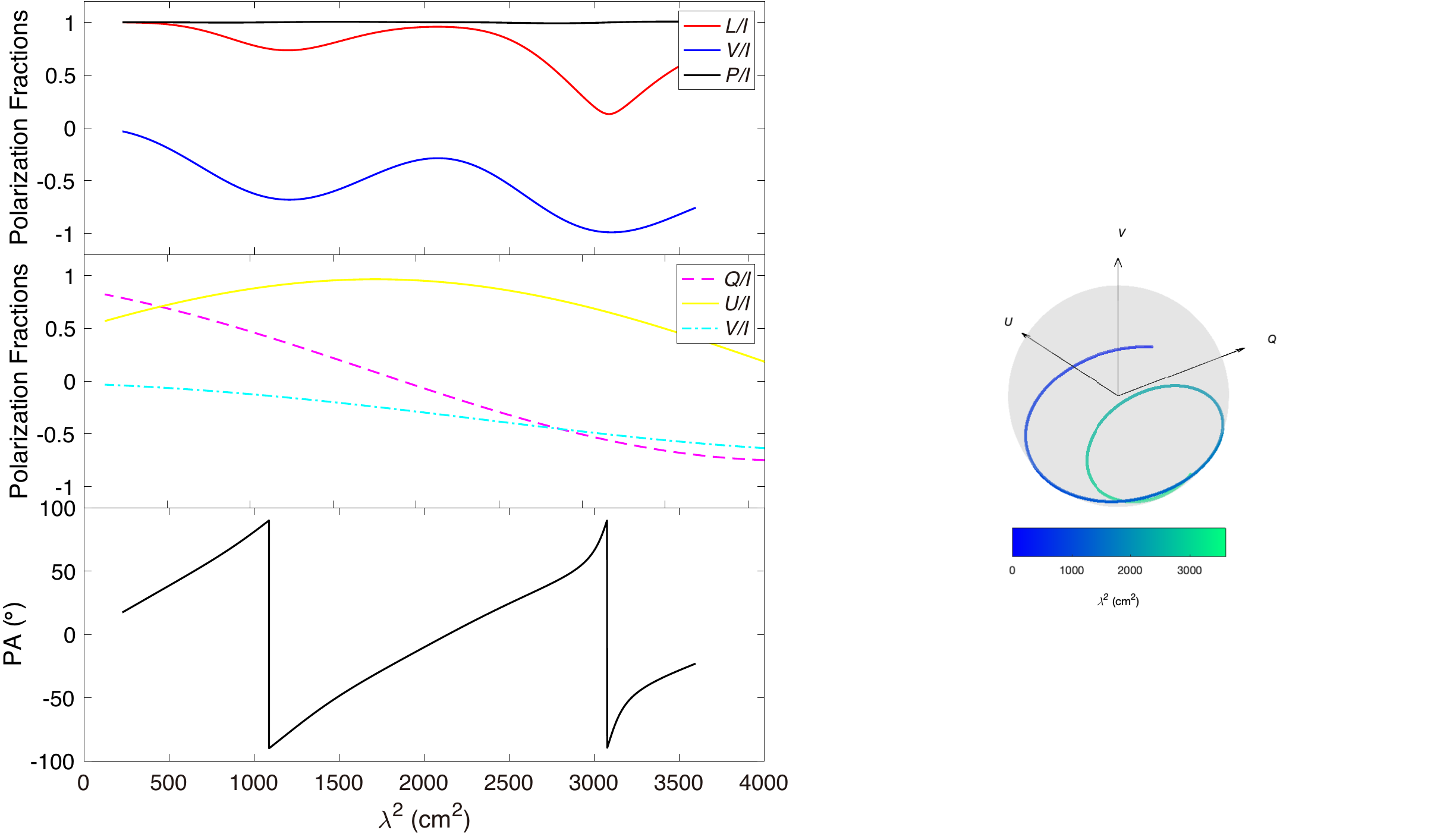}
\caption{Same as Figure~\ref{fig:prop1} but for relativistic plasma. The parameters are adopted as $\theta_B= 45^\circ$, $L =10^{14}$\,cm, $B = 1\times 10^{-3}$\,G, $K = 10^9\,\rm{cm}^{-3}$, $\gamma_i=100$, $\alpha=1.0$, $Q_0/I_0 = 1$, $U_0/I_0 = 0$, and $V_0/I_0 = 0$
\label{fig:prop2}}
\end{figure*}

For a magnetized relativistic plasma, the energy distribution could be non-thermal, and the plasma emits strong synchrotron radiation.
We consider that the plasma is characterized by a power law distribution, e.g., $dn/(d\gamma d\Omega_\psi)=K\gamma^{-s}\Theta(\gamma-\gamma_i)g(\psi)$, where $\Theta(\gamma-\gamma_i)$ is the step function.
Assuming an isotropic pitch-angle distribution ($g(\psi)=1$), the emissivity, absorption, and mixing Faraday coefficients are given by~\citep{1965ARA&A...3..297G,1969SvA....13..396S,1977ApJ...214..522J}.
\begin{equation}
\begin{aligned}
\epsilon_{I} & =\frac{3^{\alpha+1 / 2}}{\alpha+1} \Gamma\left(\frac{\alpha}{2}+\frac{11}{6}\right) \Gamma\left(\frac{\alpha}{2}+\frac{1}{6}\right)\pi mcr_e\nu_{B \perp}K\left(\frac{\nu_{B \perp} }{\nu}\right)^\alpha, \\
\epsilon_Q & =\frac{3^{\alpha+1 / 2}}{\alpha+\frac{5}{3}} \Gamma\left(\frac{\alpha}{2}+\frac{11}{6}\right) \Gamma\left(\frac{\alpha}{2}+\frac{1}{6}\right)\pi mcr_e\nu_{B \perp}K\left(\frac{\nu_{B \perp} }{\nu}\right)^\alpha, \\
\epsilon_V & =-\frac{3^\alpha}{2} \frac{\left(\alpha+\frac{3}{2}\right)}{\left(\alpha+\frac{1}{2}\right)} \Gamma\left(\frac{\alpha}{2}+\frac{11}{12}\right) \Gamma\left(\frac{\alpha}{2}+\frac{7}{12}\right) 4\pi mcr_e\nu_{B \perp}K\left(\frac{\nu_{B \perp} }{\nu}\right)^{\alpha+1 / 2} \cot \theta_B,\\
\eta_I &=\frac{3^{\alpha+1}}{4} \Gamma\left(\frac{\alpha}{2}+\frac{25}{12}\right) \Gamma\left(\frac{\alpha}{2}+\frac{5}{12}\right) 4\pi\frac{r_e c}{\nu_{B \perp}}K\left(\frac{\nu_{B \perp} }{\nu}\right)^{\alpha+5 / 2},\\ \eta_Q &=\frac{3^{\alpha+1}}{4} \frac{\left(\alpha+\frac{3}{2}\right)}{\left(\alpha+\frac{13}{6}\right)} \Gamma\left(\frac{\alpha}{2}+\frac{25}{12}\right) \Gamma\left(\frac{\alpha}{2}+\frac{5}{12}\right) 4\pi\frac{r_e c}{\nu_{B \perp}}K\left(\frac{\nu_{B \perp} }{\nu}\right)^{\alpha+5 / 2}, \\
\eta_V &=\frac{3^{\alpha+1 / 2}}{2} \frac{(\alpha+2)}{(\alpha+1)}\left(\alpha+\frac{3}{2}\right) \Gamma\left(\frac{\alpha}{2}+\frac{7}{6}\right) \Gamma\left(\frac{\alpha}{2}+\frac{5}{6}\right) 4\pi\frac{r_e c}{\nu_{B \perp}}K\left(\frac{\nu_{B \perp} }{\nu}\right)^{\alpha+3} \cot \theta_B,\\
\rho_Q&=-4\pi\frac{r_e c}{\nu_{B \perp}}K\left(\frac{\nu_{B \perp} }{\nu}\right)^3 \gamma_i{ }^{-(2 \alpha-1)}\left\{\left[1-\left(\nu_i / \nu\right)^{\alpha-1 / 2}\right]\left(\alpha-\frac{1}{2}\right)^{-1}\right\}, \quad\left(\alpha>\frac{1}{2}\right), \\
\rho_V & =8\pi \frac{\left(\alpha+\frac{3}{2}\right)}{(\alpha+1)}\frac{r_e c}{\nu_{B \perp}}K\left(\frac{\nu_{B \perp} }{\nu}\right)^2\left(\ln \gamma_i\right) \gamma_i{ }^{-2(\alpha+1)} \cot \theta_B,
\end{aligned}
\label{eq:parameters}
\end{equation}
where
\begin{equation}
  \nu_{B \perp} \equiv \frac{e B \sin \theta_B}{2 \pi m c}, \quad \nu_i \equiv \gamma_i^2 \nu_{B \perp} , \quad \alpha=\frac{s-1}{2}. 
\end{equation}

Compared with the cold plasma scenario, the Faraday rotation rate is suppressed by the same factor of $(\ln\gamma_i)\gamma_i^{-2(\alpha+1)}$, while the Faraday conversion rate is multiplied by a factor of $\gamma_i^{-(2\alpha-1)}$.
This makes it more possible that the Faraday rotation coefficient becomes comparable to that of conversion.
The Razin effect works when $\omega<\gamma \omega_p$, which makes the synchrotron spectrum be cut off due to the suppression of beaming.
This imposes a constraint of $n_0<3\times10^4\,\rm{cm}^{-3}$~\citep{2020ApJ...895....7Y}.
While among the parameters of the transfer equation, there is a degeneracy among $n_0$, $\gamma_i$, and $L$.
We let $\theta_B= 45^\circ$, $B = 1\times 10^{-3}$\,G, $K = 10^9\,\rm{cm}^{-3}$, $\gamma_i=100$, $I_0=10^{29}\,{\rm erg\,s^{-1}\,cm^{-2}\,Hz^{-1}}$ to study the polarization spectrum that has comparable Faraday rotation and conversion.
We set the index $\alpha = 1$, corresponding to a relatively steep spectrum.
If the power law index of the electron distribution is too flat, then according to Equation (\ref{eq:parameters}), the optical depth may exceed 1, which would prevent the FRB from escaping the plasma medium.
Observations of FRB 20190520B constrain the transverse size of a PRS to $<9$ pc, and a lower limit of $\gtrsim0.22$ pc can be inferred from the spectrum~\citep{2023ApJ...958L..19B,2023ApJ...959...89Z}.
Here, we take $L=10^{14}$~cm.
The simulated polarization profiles are shown in Figure~\ref{fig:prop2}.

\begin{figure}[ht!]
\plotone{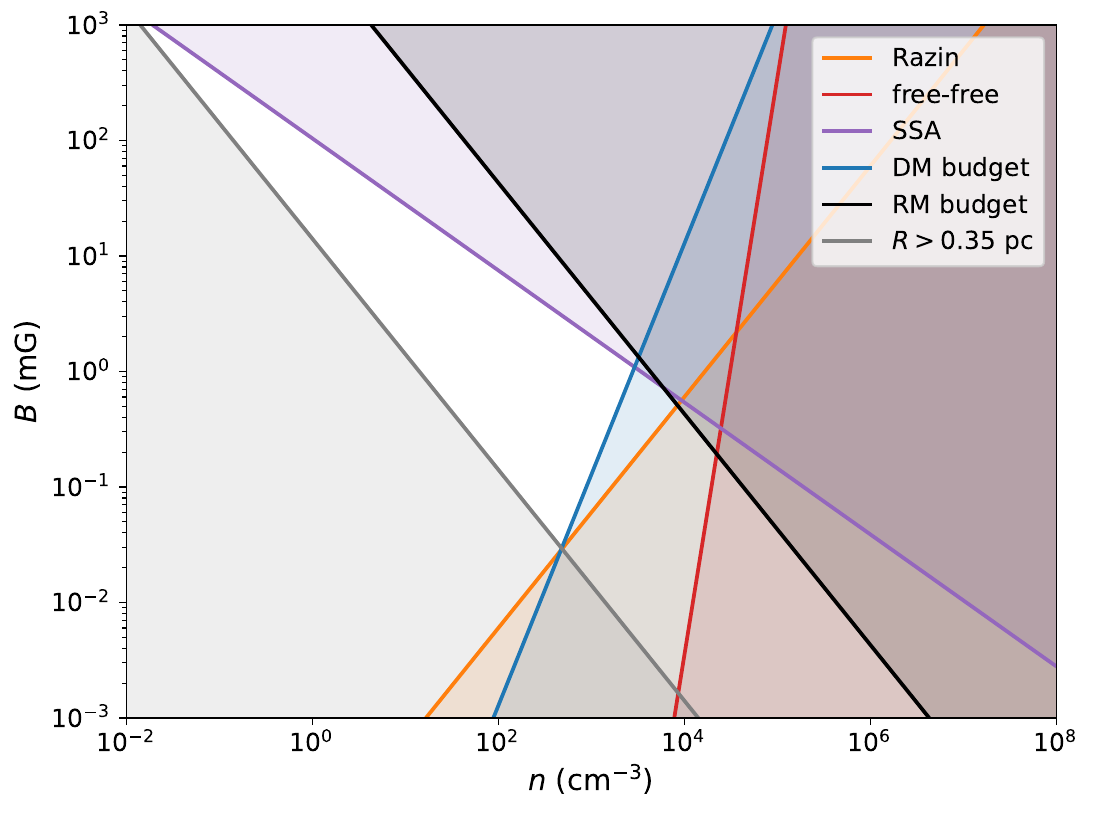}
\caption{The parameter space of the relativistic plasma medium, taking FRB 20121102A as an example.}
\label{fig:epsilon}
\end{figure}

We consider the relativistic plasma as the PRS associated with FRB.
Taking FRB 20121102A as an example, we adopt a power-law electron distribution with index $s = 2.32$, following \citet{2020ApJ...896...71L}. For each electron density $n$ and magnetic field strength $B$, we infer the radius $R(n,B)$ by matching the source-frame specific luminosity $L_\nu = 1.924 \times 10^{29} ~\mathrm{erg~s^{-1}~Hz^{-1}}$ inferred from the observed $6$~GHz flux density \citep{2017Natur.541...58C}. Interpreting the VLBI limit of $L=0.7$~pc on the projected extent as a diameter limit, we require $R(n,B) < 0.35$~pc \citep{Marcote_2017}.
The detection of the PRS at $0.4$~GHz constrains Razin suppression \citep{2021A&A...655A.102R}. For burst propagation, we require the optical depths to synchrotron self-absorption and free-free absorption to be below unity at 1 GHz.
This source has an extremely large RM value~\citep{2018Natur.553..182M}, which may also be contributed by the PRS.
Therefore, we also adopt its RM and the DM of its host galaxy as constraints.
For other FRBs, one can ignore the RM and DM constraints.
By considering the source size, the Razin effect, self-synchrotron absorption, free-free absorption, the RM and DM budgets, we plot the parameter space shown in Figure~\ref{fig:epsilon}.
It can be seen that only for specific conditions can an FRB propagate through the plasma medium while the medium itself generates sufficient power to support the PRS.
Note that we have assumed the presence of some cold plasma in the plasma medium, $\zeta\gamma^2_c= 0.01$ (e.g., \citealt{2016ApJ...819L..12Y}), which is the main contributor to the RM. Within the parameter space that ensures the FRB can propagate through the PRS, the relativistic plasma is likely to have a negligible effect on the FRB polarization.

\section{Conclusion and Discussion} \label{sec4}

Observations of FRBs now provide a rich set of polarization data, some of which exhibit frequency-dependent evolution of the Stokes parameters.
We develop an analytical solution describing the transformation of polarized emission in Stokes basis. The conclusions are summarized below.

1) We develop the analytical solution of radiative transfer of the polarization properties to a cold magnetized plasmas beyond the high-frequency approximation.
By investigating the dielectric tensor, we derive the Faraday rotation and conversion coefficients.
Both of the two coefficients are enhanced by a factor of $\omega^2/(\omega^2-\omega_B^2)$.
Compared with the case where only the zeroth-order approximation is considered, the Faraday effect here causes the Stokes parameters $Q$, $U$, and $V$ to evolve with frequency more frequently.

2) This work extends the radiative transfer solution of \cite{2025ApJ...988..164W} by invoking the emissivity.
When an FRB propagates through a relativistic plasma, it can also cause $Q$, $U$, and $V$ to evolve with frequency, but the magnetic field required in this case is smaller than that in the cold plasma scenario.
We investigate this relativistic plasma medium as a PRS and find that, although it can account for the observed PRS luminosity, an FRB can propagate through it likely under specific parameter conditions.
It would be possible to determine whether an FRB can propagate through the PRS by observing the spatial scale and luminosity of it, thereby inferring the origin of the PRS.

3) We apply the cold and highly magnetized plasma model to the study of GPM J1839$-$10 and revisit FRB 20180301A and FRB 20201124A.
The fitting results for FRB 20201124A show no significant change compared with previous results.
By considering the cold plasma model beyond the high-frequency approximation, it appears that the magnetic field of FRB 20180301A does not need to exceed 100~G (for 1 GHz).
GPM J1839$-$10 also has a similar environment.
These magnetic fields are almost perpendicular to the LOS, leading to Faraday conversion.

Unlike the GFR method~\citep{1997JPlPh..58..735M,1998PASA...15..211K}, which is a phenomenological or geometric construction assuming a rigid rotation of the polarization vector about an arbitrary diagonal axis on the Poincar\'e sphere, our analytical solution links the rotation axis and its frequency dependence directly to the underlying plasma parameters.
The mapping onto medium properties is therefore more explicit, although the parameter inversion may still be subject to degeneracies.
If the heating of the plasma medium by the FRB as well as complicated nonlinear interactions can be neglected, the solution of the transfer equation are always applicable.
Therefore, for different plasma environments, one only need to change the coefficients in the transfer equations, without having to solve for the electric field vector from scratch. 
The present analytical treatment assumes a homogeneous plasma. The effects of spatial density variations on radio-wave propagation have also been investigated numerically in unmagnetized media~\citep{2025ApJ...985..165X}.

According to Equation (\ref{eq:rho}), the mixing Faraday coefficients have two branches around $\omega=\omega_B$.
The Faraday effect arises from the differences in the dispersion relations of the wave modes.
We examine the cutoff frequencies of wave modes.
For a case where the LOS is oblique to $\boldsymbol{B}$, the cutoff frequency is given as $\omega_c^2 = \omega_p^2 + \frac{\omega_{B}^2}{2} \pm \sqrt{\frac{\omega_{B}^4}{4} + \omega_p^2 \omega_{B}^2 \cos^2 \theta}$.
If $\omega_B\gg\omega_p$, one cutoff is $\sim\omega_p^2/\omega_B$ and the other is $\sim\omega_B$.
Therefore, we focus on the results for the branch where $\omega_B<\omega$.

Cyclotron resonance occurs when the wave frequency matches the electron cyclotron frequency in the local magnetic field.
The wave energy is transferred to the electrons via cyclotron resonance, causing them to gyrate. In a cold, collisionless plasma this resonant response would be a singular (delta-function) absorption at the cyclotron frequency, which must be regularized by broadening mechanisms, such as collisions (collisional damping), thermal or relativistic broadening of the cyclotron frequency, and magnetic field inhomogeneity, before it can be interpreted as a finite absorption.
Both the cyclotron resonance and
Faraday conversion have been seen in GPM J1839$-$10~\citep{2025SciA...11P6351M}, suggesting a turbulent magnetic field which can be even up to $\sim100$~G.
A binary system with a weakly magnetized companion is a likely scenario.
In the binary system, the companion possesses only a weak magnetic field, as is typical for the late-type main-sequence stars that have been suggested as possible optical counterparts~\citep{2026SCPMA..6969512M}.

%% Please use the acknowledgment and contribution environments. This will 
%% be anonomyized when the "anonymous" style option is used. 
\begin{acknowledgments}
We are grateful to Manisha Caleb, Kejia Lee, Zenan Liu, Rui Luo, Chenhui Niu, Heng Xu, Bing Zhang and Weiwei Zhu for helpful discussions.
W.-Y. W. acknowledges support from the NSFC (No.12403058), and the Strategic Priority Research Program of the CAS (No. XDB0550300).

\end{acknowledgments}

\bibliography{sample701}{}
\bibliographystyle{aasjournalv7}

%% This command is needed to show the entire author+affiliation list when
%% the collaboration and author truncation commands are used.  It has to
%% go at the end of the manuscript.
%\allauthors

%% Include this line if you are using the \added, \replaced, \deleted
%% commands to see a summary list of all changes at the end of the article.
%\listofchanges

\end{document}